\documentclass[12pt,preprint]{aastex}
\usepackage{natbib}
\usepackage{epsfig}

\usepackage{amssymb}
\usepackage{latexsym}
\usepackage{ifthen}
\usepackage{url}

\def\lapp{\ifmmode\stackrel{<}{_{\sim}}\else$\stackrel{<}{_{\sim}}$\fi} 
\def\gapp{\ifmmode\stackrel{>}{_{\sim}}\else$\stackrel{>}{_{\sim}}$\fi}

\newcommand{\fhour}{^{\mathrm{h}}}
\newcommand{\fmin}{^{\mathrm{m}}}
\newcommand{\fsec}{\mbox{\ensuremath{.\!\!^{\mathrm{s}}}}}
\newcommand{\fdeg}{^{\circ}}

\newcommand{\psr}{J1832+0029}

\begin{document}

\title{Radio and X-ray observations of the intermittent pulsar \psr}

\author{D.R.~Lorimer$^{1,2}$, A.G.~Lyne$^3$, M.A.~McLaughlin$^{1,2}$, 
M.~Kramer$^{3,4}$, G.G.~Pavlov$^{5,6}$, and C.~Chang$^5$}
\altaffiltext{1}{Department of Physics, West Virginia University,
White Hall, Morgantown, WV 26506}
\altaffiltext{2}{National Radio Astronomy Observatory, Green Bank,
WV 24944}
\altaffiltext{3}{Jodrell Bank Centre for Astrophysics, The University of Manchester, Alan Turing Building, Manchester M13 9PL, UK}
\altaffiltext{4}{Max-Planck-Institut f\"ur Radioastronomie, Auf dem H\"ugel 69, 53121 Bonn, Germany}
\altaffiltext{5}{Department of Astronomy and Astrophysics, The Pennsylvania State University, 525 Davey Lab., University Park, PA 16802}
\altaffiltext{6}{St.-Petersburg State Polytechnical University, Polytekhnicheskaya ul. 29, St.-Petersburg 195251, Russia}

\begin{abstract}
We report on radio and X-ray observations of PSR~\psr, a 533-ms radio
pulsar discovered in the Parkes Multibeam Pulsar Survey. From radio
observations taken with the Parkes, Lovell and Arecibo telescopes, we
show that this pulsar exhibits two spindown states akin to PSRs
B1931+24 reported by Kramer et al. and J1841$-$0500 reported by Camilo
et al.  Unlike PSR~B1931+24, which switches between ``on'' and ``off''
states on a 30--40 day time-scale, PSR~\psr\ is similar to
PSR~J1841$-$0500 in that it spends a much longer period of time in the
off-state. So far, we have fully sampled two off-states. The first one lasted
between 560 and 640 days and the second one lasted between 810 and 835 days.
From our radio
timing observations, the ratio of on/off spindown rates is $1.77 \pm
0.03$.  {\it Chandra} observations carried out during both the on-
and off-states of this pulsar failed to detect any emission. Our
results challenge but do not rule out models involving accretion onto
the neutron star from a low-mass stellar companion. In spite of
the small number of intermittent pulsars currently known, difficulties
in discovering them and in quantifying their behavior imply that
their total population could be substantial.
\end{abstract}

\keywords{pulsars: individual (PSR~B1931+24; PSR~\psr; PSR~J1841$-$0500)}

\section{Introduction}\label{sec:intro}\setcounter{footnote}{0}

It is well established that not all radio pulsars emit radiation
during each rotation. Backer (1970) \nocite{bac70} first observed this
phenomenon and demonstrated that some pulsars exist in a ``null
state'' for several pulse periods before switching back on
again. Pulsar nulling has been investigated extensively over
the years (e.g., Ritchings 1976; Rankin 1986; Biggs 1992).
\nocite{rit76,ran86,big92a} From a study of pulsars in the Parkes
Multibeam Pulsar Survey (PMPS; Manchester \nocite{mlc+01} et
al.~2001), \nocite{wmj07} Wang et al.~(2007) confirmed earlier
evidence (Ritchings 1976) that
the fraction of nulling pulses generally increases with increasing
characteristic age.

In addition to the nulling phenomenon, it has become apparent that a
new class of intermittent pulsars exist where no radiation is observed
over much longer time-scales.  In a single-pulse analysis of archival
PMPS data, McLaughlin et al.~(2006) \nocite{mll+06} discovered a new
class of neutron stars (Rotating Radio Transients) from which
radio emission is detectable, on average, only 1~s per day in an
apparently random fashion. In the same year, Kramer et
al.~(2006)\nocite{klo+06} reported the discovery of a more
deterministic type of intermittency in PSR~B1931+24, which appears to
be the prototype of a large population of pulsars that have so
far been difficult to detect. As Kramer et al.~demonstrated,
PSR~B1931+24 shows a quasi-periodic on/off cycle with a period of
30--40 days in which the spindown rate increases by $\sim 50$\% when
the pulsar is in its on-state compared to the off-state. In this paper,
we report observations characterizing intermittent
behavior in PSR~\psr\ an apparently ordinary 533-ms pulsar with a
characteristic age of 5.6~Myr which was discovered as part of the PMPS
(Lorimer et al.~2006). \nocite{lfl+06} Earlier accounts of this work
we presented by Kramer (2008) and Lyne
(2009).  \nocite{kra08,lyn09} Very recently, Camilo et al.~(2012)
\nocite{crc+12} announced the discovery of PSR~J1841$-$0500, a 912-ms
pulsar which has so far shown one off-state lasting 580~days. Like
B1931+24, the spindown rate in the on-state is higher than the
off-state. For J1841$-$0500 the increase is approximately 150\%!
These pulsars are dramatic examples of a newly recognized and large
group of pulsars which show changes in
their emission properties and period derivatives \citep{lhk+10}
which are correlated and often quasi-periodic.
Understanding pulsar intermittency will shed
new insights into neutron star physics and populations.

The long off-states of intermittent pulsars are in stark contrast to
the longest known quiescence times of nulling pulsars, i.e.~they
exceed the typical nulling time scale by about five orders of
magnitude. In addition, the observed increase in spindown rate points
to a significant increase in the magnetospheric particle outflow when
the pulsar switches on, indicating that a pulsar wind plays a
significant role in neutron star spin evolution. As described by
Kramer et al.~(2006), and discussed later in this paper, the spindown
rate changes allow us to estimate the current density associated with
the radio emission.

The difficulties in detecting and identifying intermittent pulsars
imply that the few we currently observe represent a potentially
substantial population of similar objects in the Galaxy. To better
understand this population, it is therefore important to
establish the related time-scales for the non-emitting state.  Here
we detail our observations of intermittent behavior in PSR~\psr. In \S
\ref{sec:Radio} we describe the radio observations we have carried out
to characterize its intermittency. In \S \ref{sec:Xray} we describe
the {\it Chandra} X-ray observations which constrain the high-energy
emission from the pulsar. In \S \ref{sec:discussion}, we discuss
the implications of our results.

\section{Radio observations of PSR~\psr}\label{sec:Radio}

PSR~\psr\ was first detected during the PMPS in an observation made on
2000 November 23.  Following its confirmation observation in 2003
September as part of follow-up observations for the survey (Lorimer et
al.~2006), \nocite{lfl+06} PSR~\psr\ has been observed regularly by
both the Parkes 64-m and Lovell 76-m radio telescopes. Parkes
observations span the period from 2004 September until 2008
September. Lovell observations began in 2006 March and continue to be
made. To date, a total of 422 individual detections of the pulsar have
been collected (331 of these with the Lovell telescope). The initial
Parkes observations provided routine detections of the pulsar at 1374
and 1518 MHz in five minute pointings until 2004 May.  Despite 27
observations of the pulsar between 2004 June 26 and 2006 January 7, the
pulsar was not detected. By 2005 March, the Parkes observation times
had increased to 20 minutes. The non-detections mean that the flux
density of PSR \psr\ must have been less than 70~$\mu$Jy during
the period 2004 May to 2005 February for the 5-minute observations and
below 40~$\mu$Jy for the 20-min observations carried out between 2005
March and 2006 January.  The pulsar was finally detected again on 2006
March 3 in a 20-min observation at 1374~MHz.  It was subsequently
routinely detectable, primarily with the Lovell telescope,
until 2010 April 7, after which it became undetectable 
until 2012 July 20 when it resumed its regular behavior.

In all of the
observations in which the pulsar was detected, we find no evidence for
any variation in the pulse width (7.1~ms full-width at half maximum)
or flux density (140~$\mu$Jy) as originally presented in Lorimer et al.~(2006).
Currently our Parkes and Lovell observations provide only modest
constraints on the degree of polarization in PSR~\psr\ and indicate
that it is less than 20\% linearly polarized. As we discuss later
(\S \ref{sec:magnetosphere}), sensitive observations
of this pulsar's polarimetric properties could be valuable in helping
to discriminate between various proposed models 

To make the most stringent constraints of PSR~\psr\ during its off-state,
on 2011 March 8 and 9 we carried out two 1-hr observations using the
305-m William E. Gordon Arecibo radio telescope. Both observations
were conducted using the L-band wide
receiver\footnote{http://www.naic.edu/$\sim$astro/RXstatus/Lwide/Lwide.shtml}
and the Wideband Arecibo Pulsar Processors (WAPPs; Dowd et
al.~2000)\nocite{dsh00}.  Four WAPPs were used to measure three-level
autocorrelation functions every 128~$\mu$s in each of four 100-MHz
sub-bands spanning 1100--1500~MHz. The data from each WAPP were
Fourier transformed using the {\tt filterbank} program within the {\sc
  sigproc} software package\footnote{http://sigproc.sourceforce.net}
to synthesize a filterbank with 512 frequency channels, each of width
781.25~kHz.  These channelized data were then dedispersed at the pulsar's dispersion
measure (DM) of 28.3 cm$^{-3}$~pc and folded over a range of trial periods around the
nominal value predicted by our timing model using {\sc sigproc}'s {\tt fold}
program. A blind periodicity
search was also carried out using {\sc sigproc}'s {\tt seek} program
down to a signal-to-noise ratio (S/N) of 6. PSR~\psr\ was not detected
in either of these analyses. Assuming an average gain of 8.5~K/Jy
which is appropriate for the large zenith angles during the
observation (17--20 degrees), these S/N limits in the searching 
and folding analyses translate to
an upper limit on the flux of 2--3~$\mu$Jy in each observation.
Folding the data coherently across both days also resulted in no
detection, with a corresponding upper limit of 1.6~$\mu$Jy. At its
nominal on-state flux density of 140~$\mu$Jy, the pulsar would have been
detected with S/N of approximately 440. We can therefore constrain any
emission in the off-state to be less than one part in 440 of the flux
density in the on-state. At the
distance $d=1.3$~kpc inferred from the pulsar's dispersion
measure and the Cordes \& Lazio (2002) \nocite{cl02} free electron
distribution model, the upper limit on the 1400-MHz luminosity
from these observations is 2.7~$\mu$Jy~kpc$^2$. This limit is
a factor of 16 lower than that inferred for the off-state of PSR B1931+24
(Kramer et al.~2006) and an order of magnitude
below the faintest pulsar currently known (PSR~J2144--3933; Manchester et 
al.~1996). \nocite{mld+96}

\section{X-ray observations of PSR~\psr}\label{sec:Xray}

PSR~\psr\ was observed with the Advanced CCD Imaging Spectrometer
(ACIS) on board {\sl Chandra} on 2007 October 19 (observation
ID 9145) and 2011 March 28 (observation ID 12256), when the pulsar was
in on- and off-states, respectively. In both observations, the
target was imaged near the optical axis on the S3 chip in Timed Exposure
mode with a frame time of 3.24~s. Other activated chips were I2, I3,
S1, S2, and S4.  The data were telemetered in Very Faint format,
optimal for distinguishing between X-ray events and events associated
with cosmic rays. The useful effective exposure times (livetimes)
were 19.87~ks and 22.75~ks for the first and second observations,
respectively.

\begin{figure}[ht]
\centering
\includegraphics[scale=0.42]{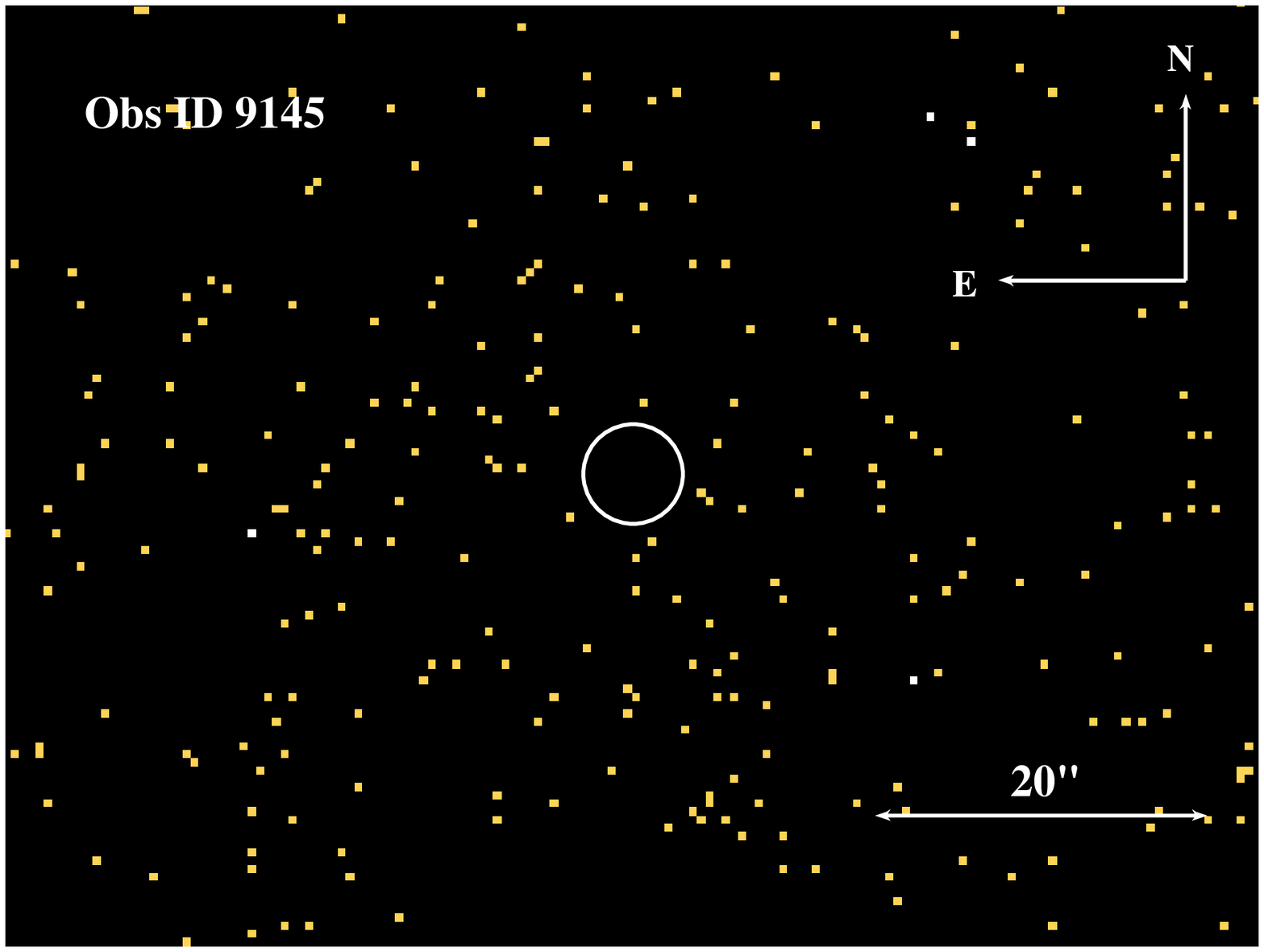}
\includegraphics[scale=0.42]{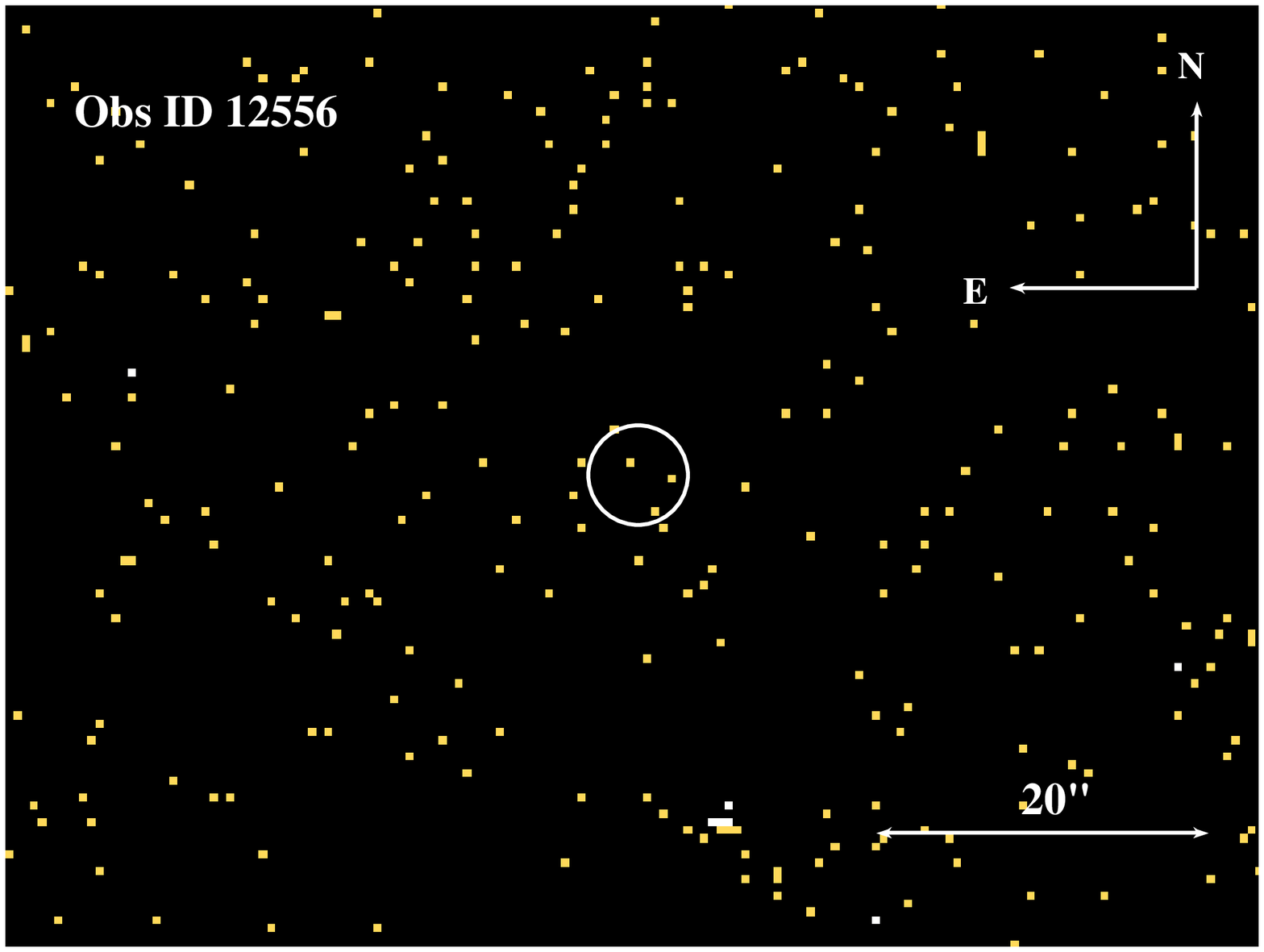}
\caption{ACIS-S3 images of the PSR\psr\ field for the
  observations of 2007 October 19 (ObsID 9145) and 2011 March 28 (ObsID 12556),
  in the 0.3--8 keV band. The images have been corrected using
  astrometric measurements of 2MASS sources (see text).  The $r=3''$
  circles are centered on the radio pulsar position from Table 1. 
  The nearest
  detected X-ray object (barely seen in the first observation) is
  located $\approx 20''$ south of the pulsar.  }
\label{fig:acis}
\end{figure}

Inspection of the ACIS images shows no X-ray sources closer than
$\approx 20''$ from the pulsar position quoted in Table 1.  
To accurately place upper
limits on the pulsar's X-ray emission, we attempted to
improve absolute astrometry by cross-correlation of the ACIS positions
of field X-ray sources with their possible counterparts in the Two
Micron All Sky Survey (2MASS) catalog, using the {\tt reproject\_aspect} 
script\footnote{http://cxc.harvard.edu/ciao/threads/reproject\_aspect}.
Because of the small number of plausible matches between the X-ray and
2MASS sources (e.g., 6 sources with $<2''$ separation, within $9'$
from the aim point), the correction turned out to be substantially
dependent on the choice of sources.  We chose four 2MASS sources with
likely X-ray counterparts on the S3 chip (separations $<1\farcs2$) for
each of the observations, which resulted in 0\farcs3 and 0\farcs2
corrections in the pulsar's position for the first and second
observations, respectively (i.e., less than the size of the ACIS
pixel, 0\farcs492).  The corrected images are shown in
Fig.~\ref{fig:acis}.

We analyzed the data in the ACIS energy range of 0.3--8 keV,
conservatively choosing $r=3''$ circles around the radio pulsar
position for the source region, which includes 99\% of all source
counts, for a typical pulsar spectrum.  We measured the background,
$N_b=1140$ and 1062 counts for the first and second observations,
respectively, in the $80''$ radius circle around the pulsar position,
excluding the $r=10''$ circle around the pulsar and $r=5''$ circle
around the faint (and variable) field source $\approx 20''$ south of
the pulsar (i.e., the area of background aperture is $A_b=19,713$
arcsec$^2$).  With the background surface densities
$N_b/A_b=0.058\pm0.002$ and $0.054\pm0.002$ counts arcsec$^{-2}$, we
expect $n_b=1.64\pm 0.05$ and $1.54\pm 0.05$ background counts in the
$r=3''$ source apertures for the first and second observations,
respectively.

The detected numbers of source and background counts in the
source aperture are $n=0$ and 3, for the first and second
observations, respectively (the same as in the uncorrected
images). This means that the pulsar was not detected in our
observations, and we can only put upper limits on its count rate and
flux. Assuming Poisson statistics,
the upper limit $n_u$ on the total number of counts in the
detection area at confidence level $C$ can be estimated 
\citep[see, e.g.,][]{geh86} as follows:
\begin{equation}
C = 1 - \sum_{m=0}^{n} \frac{n_{u}^m \exp(-n_u)}{m!},
\label{ul}
\end{equation}
which leads to the upper limit $n_{s,u}=n_{u}
-n_b$ on the number of source counts.  
For the first observation ($n=0$), $n_{u}=-\ln
(1-C)$ and $n_{s,u} = 2.97$ counts for $C=0.99$.
For the second observation ($n=3$), the
corresponding upper limit for $C=0.99$ is 10.05 counts 
\citep[see Table 1 in][]{geh86}, and $n_{s,u}=8.51$ counts.
The source count rate upper limits at the 99\% confidence level
are therefore
$R_{s,u}= 1.5 \times 10^{-4}$ counts s$^{-1}$
for the first observation and $R_{s,u}= 3.7 \times
10^{-4}$ counts s$^{-1}$ for the second observation.

\section{Discussion}\label{sec:discussion}

The radio observations reported above suggest long-term intermittent
behavior in PSR~\psr, while the X-ray observations failed to detect
any difference in the high-energy emission between the on- and off-states.  
We now discuss the implications of these observations.

\subsection{Spindown behavior in the two states}

\begin{figure}[t]
\includegraphics[angle=270,width=8.5cm]{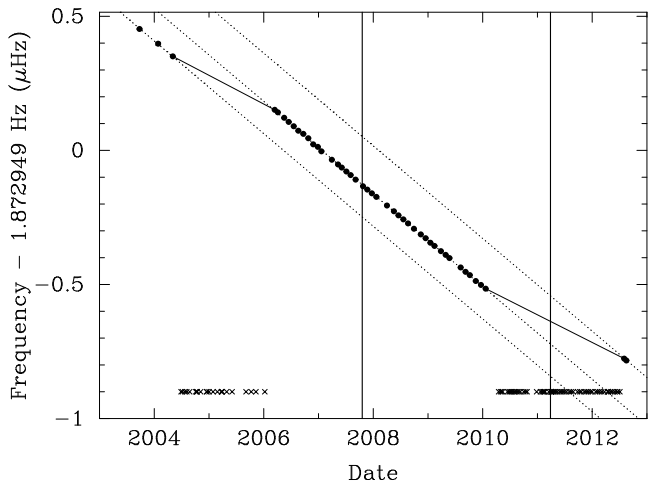}
\caption{The rotational frequency history of PSR~\psr.
The errors in the measurement of the data points are smaller
than the size of the symbols. The three dotted lines show the
$\dot{\nu}_{\rm on}$ behavior as inferred from the {\sc tempo2} fit
to the second on-state. The slanted solid lines show the frequency
derivative inferred during the off-state.
Epochs of non-detections in the Parkes
and Jodrell timing campaigns are shown by the crosses. The vertical
lines show the epochs of the two {\it Chandra} ACIS observations.}
\label{fig:spindown}
\end{figure}

To track the variation in spin frequency ($\nu$)
of PSR~\psr\ we used
the {\sc tempo2} software package (Hobbs et al.~2006)
\nocite{hem06} and its {\tt stridefit} plugin to carry out
measurements of $\nu$ based on timing model fits to short (30-day time
span) segments of data in which the position was fixed at the nominal
values found by Lorimer et al.~(2006) and the spin frequency
derivative $\dot{\nu}$ was assumed to
be zero. As shown in Fig.~\ref{fig:spindown}, this analysis reveals a
clear discontinuity in the spindown behavior during the off-states.
Two explanations could account for this: (i) the pulsar suffered
a period glitch during the off-states; (ii) the spindown rate was different
during the off-states, as is observed for PSRs~B1931+24 and
J1841$-$0500. Although it is possible to fit across the 2004--2005
period, resulting in $\Delta \nu/\nu=(5.34 \pm 0.07) \times 10^{-8}$
for the putative glitch, no exponential recovery is observed and
the abrupt turn-off observed in emission is
inconsistent with other observations of glitching pulsars \citep[see,
e.g.,][]{elsk11}. A similar conclusion can be reached for the 
2010--2012 off-state. Henceforth we examine the
hypothesis where the spindown rate of PSR~\psr\ has two distinct
values which we refer to as $\dot{\nu}_{\rm on}$ and
$\dot{\nu}_{\rm off}$.

\begin{figure}[t]
\includegraphics[angle=270,width=8.5cm]{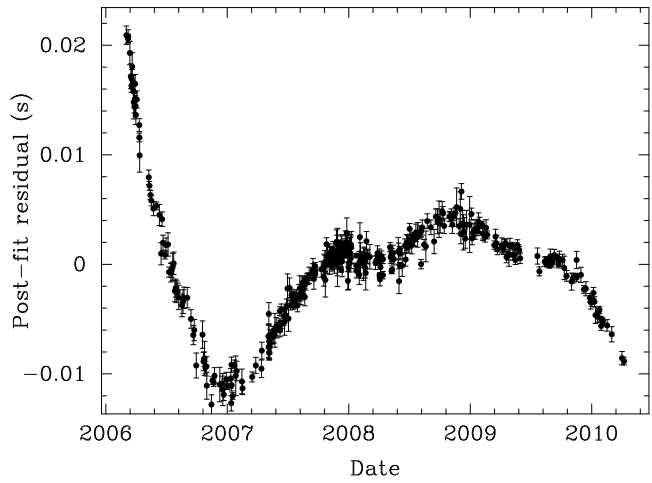}
\caption{Timing model residuals for PSR~\psr\ obtained from our
{\sc tempo2} analysis of the second on-state using the ephemeris
quoted in Table 1. The root-mean-square of the data shown here is 4.37~ms.
These data, and all other pulse arrival times collected so far are
freely available at http://astro.phys.wvu.edu/\psr.
}
\label{fig:residuals}
\end{figure}

Using {\sc tempo2}, we obtained independent timing solutions in each
of the two on-states. The results of these analyses are summarized in
Table 1. The pulsar's DM was not constrained by these analyses and was
therefore held fixed at the value reported by Lorimer et al.~(2006; DM
= 28.3~cm$^{-3}$~pc).  The shorter timing baseline ($\sim 270$~days)
for the first on-state compared to the second one ($\sim 4$~yr) means
that the timing parameters obtained from it are less precise and
subject to covariances. So far, we have only sampled $\sim 1$~month in the
current (third) on-state. To minimize these covariances, 
we held the position in the
first on-state fit fixed at the position derived from the second
on-state. While the post-fit residuals for the first on-state are
approximately white, a
significant amount of timing noise is present in the residuals from
the second on-state shown in Fig.~\ref{fig:residuals}.
This behavior can be removed by fitting
multiple sinusoids to the data (the ``harmonic whitening'' technique;
\nocite{hlk+04} Hobbs et al.~2004). To check the effect on the
measured parameters in Table 1, we carried out such an analysis using
the {\tt fitwaves} plugin to {\sc tempo2}. The residuals can be
whitened by removing six harmonically related sinusoids and the result
fit parameters are all within 1-$\sigma$ of the values presented in
Table 1. We therefore adopt the parameters from the second on-state as
being our most precise measurements of the pulsar to date and, hence,
$\dot{\nu}_{\rm on}=-(5.44505 \pm 0.00007) \times
10^{-15}$~s$^{-2}$.  

To measure $\dot{\nu}_{\rm off}$, we accounted for the uncertainties
in off/on switching epochs in the following way. We first assumed that
the nominal switch-off epoch ($T_{\rm off}$) of the pulsar occured
midway between the date of the last detection during the second
on-phase ($T_1 =$~MJD 55293.33) and the first non-detection ($T_2 =$~MJD
53301.99). Similarly, for the nominal switch-on epoch ($T_{\rm on}$), we
adopt the midpoint between the last non-detection ($T_3=$~MJD 56110.85)
and the first re-detection of the third on-phase ($T_4=$~MJD 56128.78).
Using {\sc tempo2}, we computed $\nu(T_{\rm off})$ and $\nu(T_{\rm
  on})$ --- the nominal pulse frequencies at both $T_{\rm off}$ and
$T_{\rm on}$ as predicted by the second and third on-state timing
models, respectively.  The off-state spindown rate
\begin{displaymath}
\dot{\nu}_{\rm off} = \frac{\nu(T_{\rm on})-
\nu(T_{\rm off})}{T_{\rm on}-T_{\rm off}} = -(3.08 \pm 0.05) 
\times 10^{-15}\,{\rm s}^{-2}.
\end{displaymath}
Here the uncertainty in $\dot{\nu}_{\rm off}$
is dominated by the uncertainty in $T_{\rm on}-T_{\rm off}$ which 
we estimated to be the mean of the two time windows of interest
here (i.e. $(T_4-T_3+T_2-T_1)/2)$). 
A similar analysis for the first off-state yields
$\dot{\nu}_{\rm off} = -(3.2 \pm 0.2) \times 10^{-15}$~s$^{-2}$.
These results imply that
the ratio of on/off spindown rates
${\cal R}=\dot{\nu}_{\rm on}/\dot{\nu}_{\rm off}$ is therefore 
$1.77 \pm 0.03$, i.e.~slightly higher than PSR~B1931+24 but
below PSR~J1841$-$0500.

\subsection{Charge density in the pulsar magnetosphere} 
\label{sec:magnetosphere}

To estimate the implied current flow in the pulsar
magnetosphere for both the on and off-states, we 
follow Kramer et al.~(2006) and 
consider the simplest possible emission model. We
assume that, in the off-state, the pulsar spins down by a mechanism
that does not involve a substantial particle ejection (e.g., it
would be magnetic dipole radiation if the pulsar were in vacuum), while
the rate in the on-state
is enhanced by a torque from the current of an additional plasma
outflow.  Assuming that the spindown energy loss rate in the on-state,
$\dot{E}_{\rm on}$, may be written as the sum of the energy loss rate
in the off-state, $\dot{E}_{\rm off}$, and the energy loss due to the
additional plasma, $\dot{E}_{\rm plasma}$, the corresponding charge
density (in cgs units) in the plasma 
\begin{equation}
\rho_{\rm plasma} = \frac{3 I
(\dot{\nu}_{\rm off} - \dot{\nu}_{\rm on})}{R^4_{\rm pc} B_{\rm off}}.
\end{equation}
Here $I$ is the moment of inertia of the neutron star, 
\begin{equation}
R_{\rm pc} = 
\sqrt{\frac{2 \pi \nu R^3_{\rm NS}}{c}}
\end{equation}
is the polar cap radius, $B_{\rm off}$ 
is the dipole surface magnetic field strength calculated 
from the spin frequency and spindown rate in the off-state, and
$c$ is the speed of light. For
a canonical neutron star of radius
$R_{\rm NS}=10^6$~cm and moment of inertia $I=10^{45}$~gm~cm$^2$, 
we find $\rho_{\rm plasma} \simeq
62$~esu~cm$^{-3}$.  This is slightly
higher than the so-called 
Goldreich-Julian density \nocite{gj69} 
\begin{equation}
\rho_{\rm GJ}
= \frac{B_{\rm off}}{Pc} \simeq 44 \,{\rm esu}\,{\rm cm}^{-3}
\end{equation}
which is
the charge density required to radiate along the open magnetic field lines
in the idealized pulsar magnetosphere model proposed by
Goldreich \& Julian (1969).

For PSR~B1931+24, Kramer et al.~(2006) found that $\rho_{\rm GJ}
\simeq \rho_{\rm plasma} \simeq 100$~esu~cm$^{-3}$. Taking the
corresponding values for PSR~J1841$-$0500 from Camilo et al.~(2012),
we find for this pulsar that the plasma density $\rho_{\rm plasma}
\simeq 400$~esu~cm$^{-3}$ is significantly larger than $\rho_{\rm GJ}
\simeq 130$~esu~cm$^{-3}$. The fact that these inferred densities all
equal or exceed $\rho_{\rm GJ}$ at least implies that the basic
conditions for radiation by the Goldreich \& Julian (1969) model are
being met. 

For simplicity, the above calculations make the assumption
that the pulsar is an orthogonal rotator. In reality, of course, the
inclination angle between the spin and magnetic axes $\alpha <
90^{\circ}$.  More recent and realistic modeling of the pulsar
magnetosphere by Li et al.~(2012) and Kalapotharakos et al.~(2012)
\nocite{lst12,kkhc12} consider force-free electrodynamic and resistive
solutions which can account for the different spin-down rates observed
in these three intermittent pulsars. Based on our measurement of the
on/off spindown ratio, ${\cal R}$, and the results presented in Fig.~3 of Li
et al.~(2012), the prediction for PSR~\psr\ is that $\alpha \sim
60^{\circ}$. A future Arecibo observing campaign on this pulsar during
its next on-state will be undertaken to obtain high-quality
polarimetric data with the aim of constraining $\alpha$.  As discussed by
Beskin \& Nokhrina (2007) \nocite{bn07} and Kalapotharakos et
al.~(2012), further discoveries of intermittent pulsars with different
values of ${\cal R}$ than observed so far would greatly help to constrain
these models.

In the context of models for pulsar intermittency involving the
neutron star's emission mechanism, it should also be noted that
Zhang, Gil \& Dyks~(2007) \nocite{zgd07} proposed that intermittent
pulsars are old isolated neutron stars which have entered the
so-called ``death valley'' in the $P-\dot{P}$ diagram (Chen \&
Ruderman 1993) \nocite{cr93} where the voltage across the neutron star
polar cap is no longer sufficient for pair production in the neutron
star magnetosphere, and the radio emission becomes sporadic. Zhang et
al.\ suggest that non-dipolar magnetic field configurations, similar
to the sunspot phenomenon, may be effective in such neutron stars and
temporarily rejuvenate their radio emission. It is not clear, however, 
how the quasi-periodic nature seen in PSR~B1931+24 can be explained
quantitatively in this scenario, or indeed whether it applies to
PSR~\psr\ or PSR~J1841$-$0500, since \citep[as noted by][]{crc+12}
none of these pulsars are in the death valley region.

\subsection{Constraints from the X-ray non-detections} \label{sec:xraycons}

The X-ray count rate upper limits can be used to estimate upper limits
on energy flux, which, however, depend on assumed spectrum. We know
from observations of old pulsars that their X-ray spectra can be
approximated by an absorbed power-law model with a photon index
$\Gamma \approx 2$--4 (e.g., Kargaltsev et al.\ 2006; Pavlov et
al.\ 2009). \nocite{kpg06,pkwg09} The hydrogen
column density toward the pulsar, $N_{\rm H} \approx 1 \times 10^{21}$
cm$^{-2}$, can be estimated from the DM assuming a 10\% average degree
of ionization of the interstellar medium. Using the {\sl Chandra} 
PIMMS tool\footnote{http://asc.harvard.edu/toolkit/pimms.jsp}, 
we obtain the following absorbed and unabsorbed energy fluxes for a
given source count rate $R_s$ in the first observation:
$F_{0.3-8\,{\rm keV}}^{\rm abs} = 0.63$, 0.47 and 0.48,
$F_{0.3-8\,{\rm keV}}^{\rm unabs} = 0.82$, 0.90 and 1.33, in units of
$10^{-15} (R_{s}/10^{-4}\,{\rm counts}~{\rm s}^{-1})$~erg~cm$^{-2}$ s$^{-1}$, 
for $\Gamma =2$, 3 and 4, respectively. For the second observation, the
corresponding fluxes are $F_{0.3-8\,{\rm keV}}^{\rm abs} = 0.70$, 0.55
and 0.59, $F_{0.3-8\,{\rm keV}}^{\rm unabs} = 0.91$, 1.04 and 1.64, in
the same units. Note that the same count rates correspond to higher
fluxes in the second observation because the ACIS effective area
became smaller.  Using these relations and the count rate upper limits
estimated above, we can estimate the flux upper limits at a given
confidence level.  For instance, for $\Gamma=3$ and $C=0.99$, we
obtain $F_{0.3-8\,{\rm keV}}^{\rm abs} < 0.7$ and $<2.0$,
$F_{0.3-8\,{\rm keV}}^{\rm unabs} < 1.3$ and $<3.8$, for the first and
second observations, respectively, in units of $10^{-15}$ erg
cm$^{-2}$ s$^{-1}$.  From these upper limits, one can estimate upper
limits on X-ray luminosity, $L_X = 4\pi d^2 F_X^{\rm unabs}$ and
efficiency, $\eta_X = L_X/\dot{E}$. For the on-state
(where $\dot{E} = 4.0\times 10^{32}$ erg s$^{-1}$), assuming $\Gamma=3$, we
obtain $L_{\rm 0.3-8\,keV} < 2.6\times 10^{29} (d/1.3\,{\rm kpc})^2$
erg~s$^{-1}$, $\eta_{\rm 0.3-8\,keV} < 6.5\times 10^{-4} (d/1.3\,{\rm
kpc})^2$, at $C=0.99$. Making the same assumptions for the off-state
(where $\dot{E} = 2.4\times 10^{32}$ erg s$^{-1}$), 
we obtain $L_{\rm 0.3-8\,keV} < 7.7\times 10^{29} (d/1.3\,{\rm kpc})^2$
erg~s$^{-1}$, $\eta_{\rm 0.3-8\,keV} < 3.0\times 10^{-3} (d/1.3\,{\rm
kpc})^2$, at $C=0.99$. These limits are consistent with the non-thermal
efficiencies observer in other non-recycled pulsars \citep{pccm02,zp04}.

\subsection{Implications for other intermittency models}

Our discussion so far has focused on pulsar intermittency as being due
to processes that are internal to the neutron star magnetosphere.  At
least two alternative scenarios, which we discuss below, have been
made to explain the phenomenon as being due to the influence of
material emanating from outside the magnetosphere.

Cordes \& Shannon (2008) \nocite{cs08} investigated the consequences
of debris disks around neutron stars, i.e.~metal-rich leftover
material from the supernova explosion that has aggregated into a disk
of circumpulsar material. They propose a scenario in which the
behavior seen in PSR~B1931+24 is produced by an asteroid in an
eccentric 40-day orbit which deflects material from the debris disk
into the neutron star magnetosphere. Such a process could temporarily
halt the electron-positron pair production thought to be responsible
for the radio emission.  Unfortunately, the infall rates required by
this model translate to completely undetectable X-ray fluxes.  In
addition, given that sufficiently 
high-precision timing is not possible for 
pulsars such as PSR~\psr, any periodic signatures from such small
bodies would not be detectable in its timing residuals
(Fig.~\ref{fig:residuals}).

Rea et al.\ (2008) \nocite{rks+08} suggested that accretion onto the
neutron star from a low-mass stellar companion in an eccentric orbit
close to periastron could halt pair production. In this case, the
heating of the infalling matter would produce additional X-rays in the
off-state. While no signatures indicative of a binary companion exist
in our radio timing residuals, such an orbit would not be detectable
if it were close to face-on. Rea et al.\ attempted to test this
hypothesis for PSR~B1931+24 via a {\it Chandra} ACIS observation in 2006.
Unfortunately, the pulsar switched on unexpectedly before their
observations.  To test this model in our observations of PSR~\psr,
following the discussion in \S 5.2 from Rea et al.\ (2008), we assume
that the radio emission is quenched when the neutron star's Alfven
radius is less than its light cylinder radius. This corresponds to
$L_X \gapp 10^{30}$~erg~s$^{-1}$ and is right at the boundary of
detectability in our off-state observation given the upper limit
$L_{\rm 0.3-8\,keV} < 7.7\times 10^{29} (d/1.3\,{\rm kpc})^2$
erg~s$^{-1}$ found in \S \ref{sec:xraycons}. However, since the
distance estimate to PSR~\psr\ made using the Cordes \& Lazio (2002)
electron density model can be uncertain by factors of two or more
\citep[see, e.g.,][]{dtbr09} we cannot therefore
conclusively reject this scenario as an explanation for the behavior
observed in PSR~\psr.

\subsection{How common are intermittent pulsars?}

Regardless of the form of the mechanism for pulsar intermittency, its
recognition and characterization through the three pulsars so far
poses interesting questions as to the size of the likely population of
similar objects in the Galaxy. From our sampling of PSR~\psr\ so far,
it appears to spend approximately 50\% of the time in the off-state.
Similar considerations for PSRs~B1931+24 and J1841$-$0500 imply
similar off-state duty cycles. Due to these and similar pulsars being
less likely to be on during pulsar search and confirmation
observations, as noted by Kramer et al.~(2006), they could represent a
substantial population that has so far evaded detection.
PSR~J1841$-$0500, for example, was in an off-state during the closest
PMPS observation, and was only discovered serendipitously during a
search of a magnetar in the same telescope beam \citep{crc+12}.  
In addition to
intermittent pulsars evading discovery in large-scale surveys, since a
significant fraction of the $\sim$ 1700 non-recycled pulsars currently
known are not subject to long-term timing programs, it is currently
unclear as to what fraction of these could exhibit intermittency on
long time scales. Even the prototypical object B1931+24 evaded
characterization for almost 20 years after its discovery
\nocite{stwd85} (Stokes et al.~1985). For PSR~\psr, had the initial
Parkes timing observations spanned a period of a year and no off-state
seen, it is possible that the intermittent behavior would have evaded
detection.  Perhaps the majority of normal pulsars exhibit some form
of intermittent behavior, if they are studied long enough.  If
that is the case, then current estimates of the pulsar birthrate
would need to be upwardly revised by a factor of two. The impact
of the discovery of intermittent pulsars on our understanding of the
neutron star birth rate \citep[see, e.g.,][]{kk08} is a currently
unsolved problem which merits further investigation.

\bigskip
The Parkes radio telescope is part of the Australia Telescope, which
is funded by the Commonwealth of Australia for operation as a National
Facility managed by CSIRO.  The Arecibo Observatory is operated by SRI
International under a cooperative agreement with the National Science
Foundation (AST-1100968), and in alliance with the Ana G.
M\'endez-Universidad Metropolitana, and the Universities Space
Research Association.  We thank Arun Venkataraman for retrieving the
Arecibo observations.  D.R.L., G.G.P., C.C., and M.A.M. were supported
by {\it Chandra} grants GO8-9078 and GO1-12078 during part of this work.
D.R.L. and M.A.M. were also supported by a WVEPSCoR Research Challenge
Grant and acknowledge the support provided by the Australia Telescope
National Facility distinguished visitor program during completion of
this work. The work of G.G.P. was also partly supported by NASA grant
NNX09AC84G and by the Ministry of Education and Science of Russian
Federation (contract 11.G34.31.0001). We thank Fernando Camilo and
Andrew Seymour for useful comments on the manuscript, as well as
Christine Jordan for assistance with the Lovell timing data and George
Hobbs for useful discussions concerning {\sc tempo2}.

\begin{deluxetable}{lc}
\tablecaption{Observed and derived parameters for PSR~\psr\label{tab:timing}}
\tablecolumns{2}
\tablehead{\multicolumn{2}{c}{Timing model parameters for first ``on'' period}}
\startdata
 Data span (MJD)\dotfill &     
    $52886-53152$ \\
 Rotation frequency, $\nu$ (Hz)\dotfill &     
    $1.87294940061(4)$  \\
 Frequency derivative, $\dot{\nu}$ ($10^{-15}$~s$^{-2}$)\dotfill & 
    $-5.42(1)$  \\
 Reference epoch (MJD)\dotfill &     
    $53019$ \\
\cutinhead{Timing model parameters for second ``on'' period}
 Data span (MJD)\dotfill &     
    $53796-55293$ \\
 Right ascension, $\alpha$ (J2000)\dotfill &     
    $18\fhour32\fmin50\fsec825(1)\ $ \\
 Declination, $\delta$ (J2000)\dotfill &     
    $+00\fdeg29^\prime27.0(3)\ $ \\
 Rotation frequency, $\nu$ (Hz)\dotfill &     
    $1.87294879924(1)$  \\
 Frequency derivative, $\dot{\nu}$ ($10^{-15}$~s$^{-2}$)\dotfill & 
    $-5.44505(7)$  \\
 Reference epoch (MJD)\dotfill &     
    $54545$ \\
\cutinhead{Timing model parameters for third ``on'' period}
 Data span (MJD)\dotfill &     
    $56128-56160$ \\
 Rotation frequency, $\nu$ (Hz)\dotfill &     
    $1.8729482220(8)$  \\
 Frequency derivative, $\dot{\nu}$ ($10^{-15}$~s$^{-2}$)\dotfill & 
    $-5(2)$  \\
 Reference epoch (MJD)\dotfill &     
    $56144$ \\
\cutinhead{Derived parameters for on-state}
 Spin-down energy loss rate, $\dot{E} \propto \nu \dot{\nu}$\dotfill &
   $4.0 \times 10^{32}\,$erg~s$^{-1}$ \\
 Characteristic age, $\tau_c = \nu/(2|\dot{\nu}|)$ \dotfill &
   $5.4$~Myr \\
 Surface magnetic field, $B \propto \sqrt{|\dot{\nu}|/\nu^3}$ \dotfill &
   $9.2 \times 10^{11}\,$G 
\enddata
\tablecomments{Parentheses indicate 1-$\sigma$ uncertainties on the
last digit(s) as reported by {\sc tempo2}.}
\end{deluxetable}

\end{document}